# Design for X (DfX) in the Internet of Things (IoT)

Nicole M. Radziwill and Morgan C. Benton
James Madison University
Department of Integrated Science and Technology
MSC 4102, 701 Carrier Drive, Harrisonburg VA 22802

Corresponding author email: nicole.radziwill@gmail.com

# Design for X (DfX) in the Internet of Things (IoT)

**Abstract**: “Designing quality in” is a cornerstone of modern quality management philosophies. Design for X (DfX) techniques provide guidelines, heuristics, and metrics to ensure that a particular quality attribute exists in a design. Although hundreds of papers have been published on DfX approaches, few researchers have explored systematically applying multiple DfX in a particular problem context. As the Internet of Things (IoT) evolves, boundaries between people, computers, and objects will become less distinct, underscoring the need for more holistic design. Using mixed methods, this paper examines the utility of DfX in the emerging IoT ecosystem. We identify DfX that are applicable to IoT-related design, and find gaps that demand further research and development. The results from this study can be used to help designers and quality managers select or develop appropriate DfX to use in designing components for the Internet of Things (IoT), supporting actionable strategies for quality and customer satisfaction.

## Introduction

On December 21, 2015, SpaceX made history when it launched a rocket into low earth orbit to deliver 11 satellites, and then *returned* - in one piece - to a landing pad not far away. By *designing for reusability*, SpaceX achieved what was previously unthinkable, and demonstrated that cost-effective space travel could be within reach - a tremendously disruptive potential. (“Reusability: The Key”, 2015)

> *“If one can figure out how to effectively reuse rockets just like airplanes, the cost of access to space will be reduced by as much as a factor of a hundred. A fully reusable vehicle has never been done before. That really is the fundamental breakthrough needed to revolutionize access to space.” --Elon Musk*

Design for X (DfX) techniques focus design activities so that a particular quality attribute (or group of attributes) is emphasized. Using DfX, designers explore design goals and constraints early in the product lifecycle, and consider the ramifications of their choices all the way through obsolescence, disposal, and remanufacture. Each DfX consists of guidelines, checklists, metrics, methods, and mathematical models (Chiu & Okudan, 2010) and often also includes information about benchmarks or how to leverage best practices. Before application, DfX must be evaluated for their utility within different application domains (e.g. energy, medical devices, consumer products), and must be adapted for new technological capabilities and emerging innovations.

The Internet of Things (IoT), conceptualized in Figure 1, is one of the emerging innovations against which the corpus of DfX must be critically re-evaluated. IoT represents the convergence of several enabling technologies: embedded systems in the 1970's, radio frequency identification (RFID) in the 1980's, micro-electromechanical systems (MEMS) and wireless sensor networks in the 1990's and 2000's, and intelligent agents and social networks in the 2010's. (Xu et al., 2014) Although definitions for the IoT morph with each new research paper that is published, the persistent elements include 1) autonomous (or semi-autonomous) networked intelligent agents, that are 2) embedded in products, materials, people, or other living things, 3) deployed on the macroscale, microscale, and nanoscale (Akyildiz et al., 2015), that collectively 4) produce Big Data from which powerful, real-time insights may be drawn. For a comprehensive presentation of IoT enabling technologies across all architectural layers, refer to Al-Fuqaha et al. (2015).

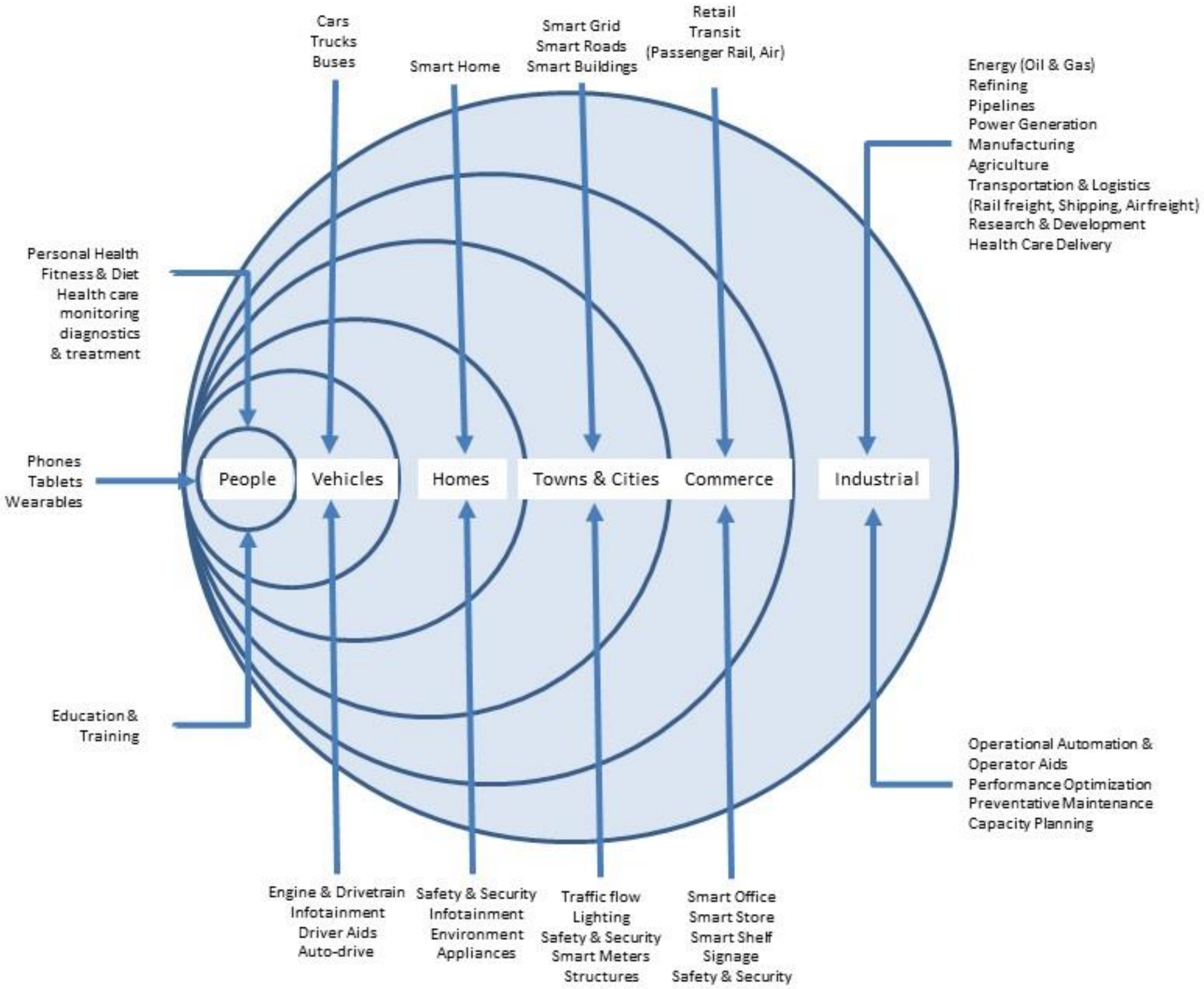

**Figure 1.** A product and service-oriented conceptualization of the Internet of Things (IoT) from http://www.iotcentral.io/blog/mapping-the-internet-of-things

This paper presents a gap analysis of DfX situated in the IoT ecosystem (which consists of a sensing layer, a communications layer, a data service layer, and an interface layer), against the backdrop of an integrative quality model that has customer satisfaction at its core. The aim of DfX is to *design quality into* a product, process, or organization - a tenet of all modern quality management philosophies. But quality is a complex and multifaceted topic. According to ISO 9001, quality is defined as the "totality of characteristics of an entity that bear upon its ability to satisfy stated and implied needs." An entity can be many things: a product, a process, a person, and organization, a resource, a project, or a complex product consisting of many assemblies and components situated within groups of people, performing tasks on their behalf. Stated needs can be obtained through research, surveys, and focus groups, but implied needs must be anticipated by the designer. Furthermore, quality is not just determined by the characteristics of a product, but by the robustness of the process that produced it, and by how the customer or stakeholder perceives value at the points of consumption, use, or engagement.

There are hundreds of thousands of articles in the scholarly literature since the 1940's that discuss quality, and nearly as many perspectives from which to discuss it. This has been one of the main challenges in research in quality since its inception. Mitra (2003) partially closed this gap by examining over 300 articles in marketing and operations management, determining how each author characterized quality, and then constructed a model to show all the passive and active definitions of quality in one place, as well as the interrelationships. Golder et al. (2012) refined and simplified the model while accentuating the importance of information flows(see Figure 2), calling out contemporaneous links (CL) or information flows that are synchronous, as well as dynamic links (DL), which are information flows that require information from a previous or subsequent time period.

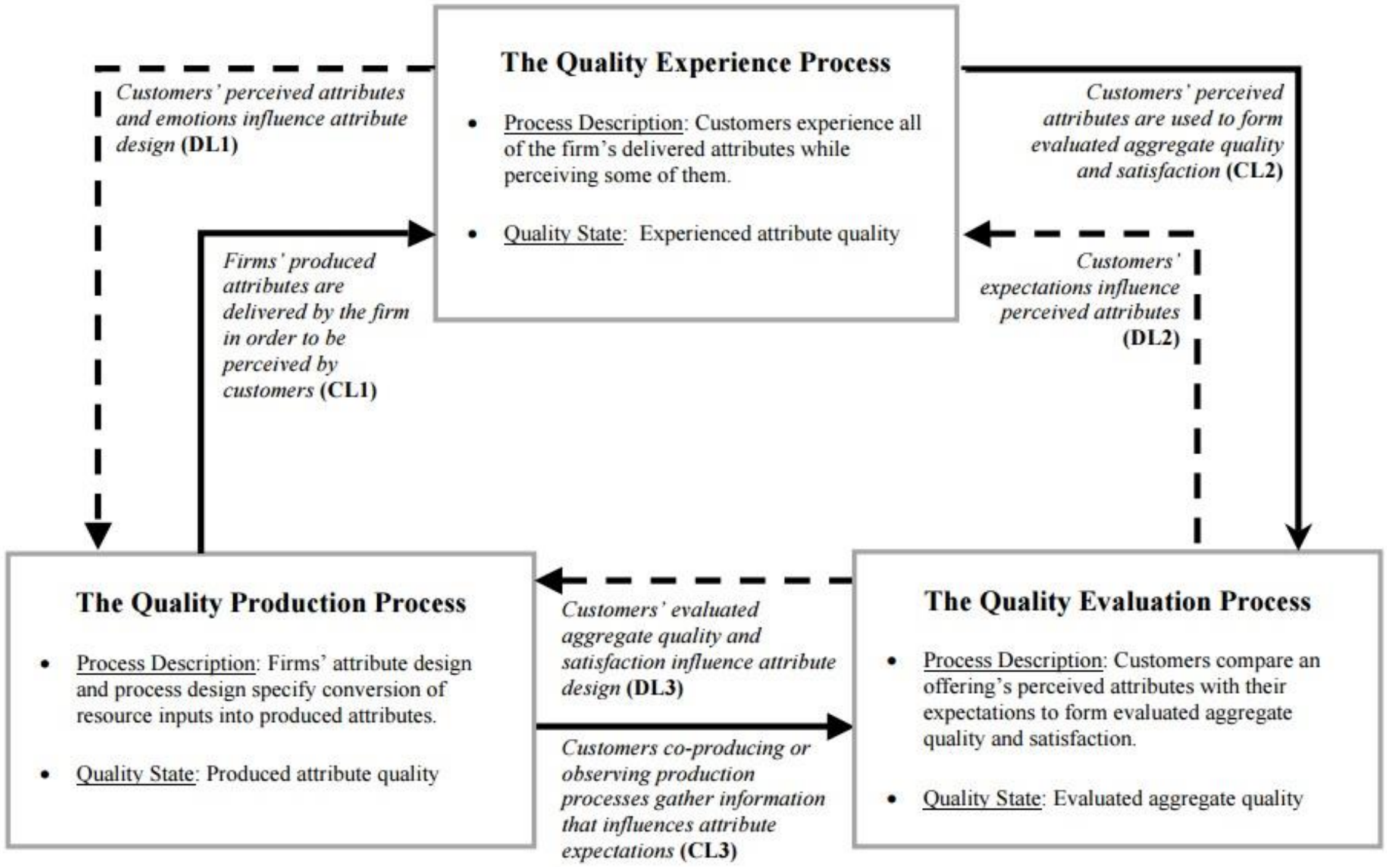

**Figure 2**. An integrative quality framework (from Golder et al., 2012)

This integrative quality framework has three main sections, within which the authors have identified several more causal links, and focuses on product quality management (even though process management is implicit). Although it was developed specifically with product lifecycle management in mind, it may also inform service quality management. Within the framework, the Quality Production Process is what most recognize as quality assurance and control. It includes process design, offline quality control methods (e.g. using simulation and operations research to achieve or exceed goals and objectives), online quality control methods (e.g. statistical process control to identify assignable cause variation), and inspection. The Quality Experience Process captures the dynamic nature of how a customer or stakeholder perceives quality and value before acquisition, during engagement, and after the product's end-of-life. The Quality Evaluation Process is the assessment and adjustment process that occurs when a customer or stakeholder's expectations regarding quality and value are met or not met.

One of the main contributions of Golder et al.'s (2012) integrative quality framework is a hierarchy of 20 actionable strategies for improving customer satisfaction. These are linked to the three sections of the framework. This hierarchy will be used later to connect DfX processes according to how firms produce and deliver quality, how customers experience quality, and how customers evaluate quality, with a focus on designing for customer satisfaction in IoT. This will ensure that the recommendations for expanding DfX for IoT are focused on quality and value generation by design.

Engineering design is the craft of building quality and value into a user's experience of a product. The newly applied technologies of the IoT will require academics and practitioners to revisit the design practice due to the omnipresence of software in objects (which, incidentally) is not new. Embedded systems (e.g. in cars, medical devices, manufacturing, and logistics systems) have been active topics in research and development for decades. Unfortunately, the research community around it has not been cohesive, so there are groups working on the technical issues of networking, communications, architecture, and IoT component development, while others envision the methodological changes that will be required to develop software for the IoT, and yet others investigate the impact of IoT on manufacturing and production. (Ebert & Salecker, 2009) What distinguishes IoT from its predecessor, embedded systems, is that the technologies are becoming more affordable and more accessible, and storage is relatively cheap.

In the "Internet of Computers" that we are all familiar with today, people produce and consume information, and use it to generate solutions (predominantly on a pull basis). In the IoT, *objects, devices, and materials* produce, consume, and aggregate the information. This generates solutions to support human judgment (via pull or push mechanisms) as well as mechanisms for the components to autonomously self-configure, self-repair, and support each other. The IoT arises as the result of a network effect: like telephones and fax machines, the emergent value that comes from recombination of information streams grows exponentially with the number of connected entities. IoT has no value as just a collection of objects with the capability to communicate. It emerges as a consequence of data and information-sharing between intelligent, connected agents, many of which will be implemented using embedded systems, and is visible through the lens of aggregated analytics. (Welbourne et al., 2009)

This study takes a first step towards integrating these research streams into a cohesive overview that will be useful for designers and quality managers. The research questions are:

- Which existing DfX processes are best suited to help designers satisfy the quality attributes that need to be designed into IoT components?
- What gaps in DfX could impact quality and customer satisfaction in the era of IoT?
- Based on these gaps, what DfX research and development is needed?

Despite the variety of directions that have been taken in the research to date, there is one common agreed upon core. In software-intensive systems engineered to be components of the IoT, the "interactions between humans and things must be mediated properly and the user should always be at the ecosystem's center," (Baresi et al., 2015) congruent with the intent of the integrative quality framework.

## Methodology

To address these research questions, this study used a convergent mixed methods approach to conduct a gap analysis on the corpus of DfX processes applied across the product lifecycle. After an extensive literature review, nonfunctional requirements for the IoT (representing quality attributes) were extracted, and merged with the list of DfX using Analytical Hierarchy Process (AHP), a quantitative tool for multi-criteria decision making. Simultaneously, the list of DfX was mapped onto the customer satisfaction strategies identified by Golder et al.'s (2012) integrative quality framework. The qualitative and quantitative results were merged via gap analysis to determine the most applicable existing DfX for IoT, and the areas where new DfX are warranted. Finally, an interpretation was generated (see Figure 3).

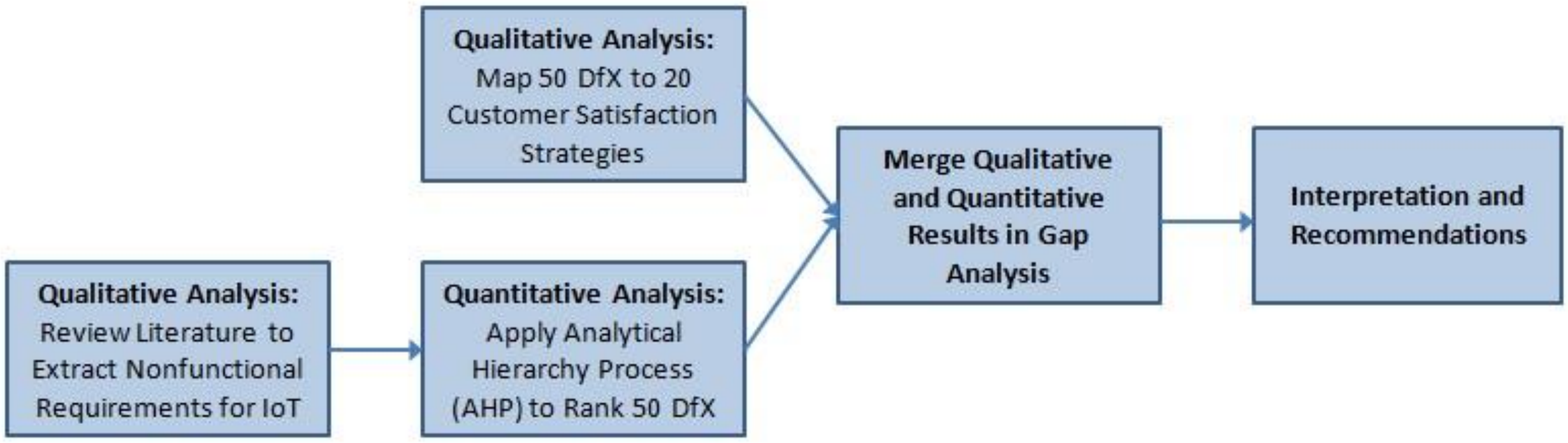

**Figure 3**. The convergent mixed methods approach used in this study.

Analytical Hierarchy Process (AHP), developed by Saaty (2008), is a quantitative method that can be used to determine which DfX best satisfy the nonfunctional requirements for the IoT. Conducting an analysis with AHP starts with defining a decision hierarchy, with a singular Goal at the top, followed by a set of Criteria at the next level, and finally a group of Alternatives to select from. Like a neural network, there can be many levels of decision criteria between the Goal and Alternatives layers. Pairwise comparisons are done between each element on a lower level, determined by which alternative best satisfies the intent of the criterion, or which sub criterion most influences the parent criterion. Priorities are determined for each level, and weighted until the priorities of all the alternatives on the lowest level can be compared to one another. One decision matrix is set up for each group of children and their parent using the rankings in Figure 4, which explains how the children compare with respect to the parent.

**Figure 4. Pairwise comparisons to be used in AHP decision matrices (from Saaty, 2008)**

| *Intensity of Importance* | *Definition* | *Explanation* |
|---|---|---|
| 1 | Equal Importance | Two activities contribute equally to the objective |
| 3 | Moderate Importance | Experience and judgment slightly favor one activity over another |
| 5 | Strong Importance | Experience and judgment strongly favor one activity over another |
| 7 | Very Strong or Demonstrated Importance | An activity is favored very strongly over another; its dominance demonstrated in practice |
| 9 | Extreme Importance | The evidence favoring one activity over another is of the highest possible order of affirmation |

As an example of how the comparisons are recorded in a matrix, consider the matrix of all lowest level Alternatives and one Product Criterion (say, for example, quality) in Figure 4. Imagine that Design for Six Sigma (DfX 1), Design for Assembly (DfX 2), Design for Miniaturization (DfX 3), and Design for Obsolescence (DfX 50) are being compared with respect to how well they can help us satisfy quality at the product level. The ellipsis indicates that there are many more DfX alternatives we could have included, and will in fact be included in the final analysis, but are not part of this example. Design for Six Sigma (DfX 1) is *extremely more important* (ranking: 9) than Design for Miniaturization (DfX 3) in satisfying the objective of quality, whereas Design for Miniaturization (DfX 3) and Design for Obsolescence (DfX 50) are of nearly equal importance (ranking: 1). Reciprocals indicate relationships in reverse (e.g. Design for Miniaturization is much less important than Design for Six Sigma in achieving the product criterion of quality). Although odd numbers are recommended, even numbers may be used if the judgments fall between two of the linguistic assessments.

For each decision matrix, we solve for the first principal eigenvector to obtain a vector of priorities. Using the `ahp` package in the R Statistical Software (ipub, 2015), priorities were synthesized between the levels until weightings are obtained for each of the alternatives on the lowest level. Because this will be such a complex decision matrix (with multiple product criteria, multiple data sub criteria, and 50 alternatives) the least significant product criteria and alternatives will be deselected for the final analysis. AHP results will indicate the most relevant DfX for IoT development. The most relevant DfX will be arranged according to Golder et al.'s (2012) customer satisfaction strategies, from the integrative quality framework, to identify critical gaps in DfX for IoT.

## Literature Review

We identified the first group of articles for potential inclusion in the analysis by a broad literature search on Google Scholar, followed by more targeted searches using EBSCOhost and ABI/INFORM, using various combinations of these keywords: *Design for X, design for, quality, Internet of Things, IoT, design for*, and *embedded systems*. This approach enabled us to capture descriptions of DfX techniques that did not use the DfX label (e.g. *affective design*) but were still focused on designing quality into products and processes. The time frames for each search were selected based on the emergence of terms in the literature: for example, the first mention of the term "Internet of Things" seems to be from 1999 (Ashton, 2009) even though it was not frequently used until it was featured in *Scientific American* five years later. (Gershenfeld et al, 2004) Research on embedded systems been conducted and published for decades.

After an initial cursory search, the results from the literature review will be presented in three parts, examined from the perspective of the quality manager (who is responsible for ensuring that quality is designed into a system as early and as effectively as possible):

1) a history and overview of all DfX techniques,
2) a description of the composition of DfX and the integration of DfX processes, and
3) the nonfunctional requirements of IoT systems, determined by mining research in ubiquitous computing, ambient intelligence, and smart environments.

The term *ambient intelligence* was frequently observed in the initial review. The co-occurrence of *ambient intelligence, Internet of Things,* and *design* has been steadily increasing since 2005, even though the number of occurrences of *ambient intelligence* peaked in 2009 at 10,200 references, and was down to 2,100 by the year 2015. The subset of these articles that also included the term *quality* was selected for part 3 of the literature review (Figure 5).

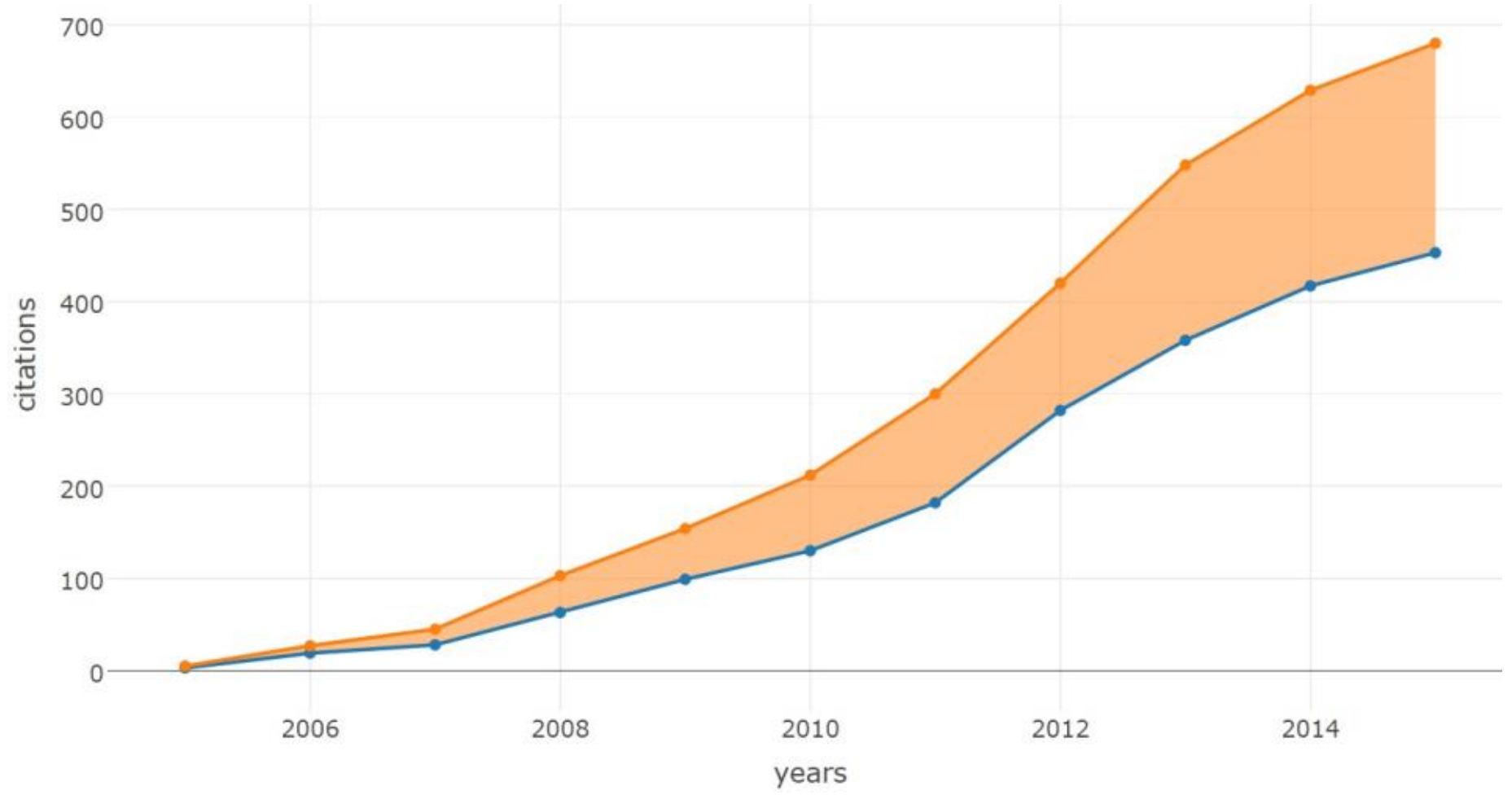

**Figure 5**. The shaded area shows the number of articles with co-occurrences of terms *ambient intelligence*, *design*, and *IoT*, with and without *quality* (the focus of part 3 of the review)

Some themes were readily apparent from filtering the initial search results of ~211,000 candidate papers. DfX appeared in 4,190 results, from which application of DfX to specific industries was excluded, with the remainder sorted by citation frequency and examined to find the first-most-cited reference for each specific DfX instance. The co-occurrence of *design, Internet of Things, and embedded systems* suggests a shift between 2000 and 2015 from purely technical research in embedded systems to the more applied research, which includes business and management context, that characterizes IoT. DfX in the context of embedded systems has focused almost solely on product lifecycle management using RFID-based product embedded information devices. (Kiritsis, 2011) The intersection of DfX and ambient intelligence was only considered in one paper, but that led to the creation of the Design for Miniaturization DfX (Niedermayer, 2007) which suggests the practical value of drawing on research in ambient intelligence.

## History and Overview of Design for X (DfX) Techniques

The phrase "Design for X" first appeared in 1990, in the proceedings for one conference (Keys, 1990) and in the *AT&T Technological Journal*. (Gatenby & Foo, 1990) Although the papers do not suggest that the authors had an awareness of each other, the tone implies that "Design for X" was not an unknown term in industry at that time. The Keys paper is the more comprehensive of the two, and explains that axiomatic design (Suh, 1990) is fundamentally compatible with DfX approaches and in fact may be the basis for many of them. Axiomatic design asserts that the design with minimal information content that preserves the independence of functional requirements is the best design. This selection process can be done by ranking the functional requirements in order of priority (using a hierarchical structure), and then conducting a rolling pairwise comparison to eliminate the designs with the greater information content. This results in selecting the design that has the greatest probability of satisfying the design goals by achieving the functional requirements. This concept is mathematically analogous to the power of a statistical test; axiomatic design, however, represents the power of a design to satisfy the functional requirements.

Search results from the initial set of references were filtered, with the goal of producing a list of all the DfX techniques discussed in the literature and their characteristics. Figure 6 classifies each of these DfX elements in terms of their: 1) purpose or goal, 2) scope, 3) character, and 4) focus. Each DfX is typically applied for a particular purpose or to achieve one or more strategic or operational goals, which were extracted from the references in Figure 4 and summarized in terms of scope, character, and focus.

The *scope* of where the DfX technique is applied can be at the product level, the system level, the ecosystem level, or a combination thereof. (Chiu & Kremer, 2001) To accommodate the wide variety of design techniques related to a particular quality attribute that were found in the literature, we expand their definition of "ecosystem" to cover not only the *natural environment* within which production and service processes are embedded, but also the *cognitive environment* and the *web of human relationships*. *Character* (from Holt & Barnes, 2010) refers to the frame of reference that a particular DfX requires: whether the development is focusing on a virtue of the product itself, or a characteristic of the system it is embedded in. *Focus* refers to the degree to which the DfX must incorporate the requirements and preferences of stakeholders outside the organization developing the product. All externally focused DfX address the needs of supply chain partners or the customer. All internally focused DfX address the specifications of the product, the requirements for the production process, or the nature of the service.

**Figure 6. DfX techniques and their characteristics.**

| *Design For* | *Purpose/Goal* | *Scope (Chiu & Kremer, 2011)* | *Character (Holt & Barnes, 2010)* | *Focus* | *References* |
|---|---|---|---|---|---|
| Cost | Reduce lifecycle costs | ALL | Virtue | Internal | Unal & Dean (1991) |
| Manufacturing (DFM) | Reduce production costs | Product | Lifecycle | Internal | Stoll (1988) |
| Assembly (DFA) | Reduce production costs | Product | Lifecycle | Internal | Boothroyd & Alting (1992); Warnecke & Babler (1988) |
| Manufacturing & Assembly (DFMA) | Reduce production costs | Product | Lifecycle | Internal | Boothroyd (1994) |
| Variety (DFV) | Reduce barriers to innovation | Product | Virtue | Internal | Martin (1999) |
| Quality * | Increase satisfaction of stated and implied needs | Product | Virtue | Both | Franceschini & Rossetto (1997) |
| Six Sigma (DFSS) * | Reduce variation and defects | Product | Both | Internal | Harry & Schroeder (2000) |
| Quality Manufacturability | Increase satisfaction of stated and implied needs; reduce production costs | Product | Both | Internal | Das et al. (2000) |
| Reusability | Reduce barriers to innovation | Product | Virtue | Internal | Cowan & Lucena (1995); Torroja et al. (1997) |
| Disassembly | Reduce environmental impact | Product | Lifecycle | Internal | Zussman et al. (1994); Zhang & Kuo (1996) |
| Reliability | Reduce failure rate | Product | Virtue | Internal | Lalli & Packard (1994); Pecht (2007) |
| Testability | Reduce failure rate | Product | Virtue | Internal | Williams & Parker (1982); Pettichord (2002) |
| Obsolescence (DfO) | Reduce supply chain costs | Product | Lifecycle | Internal | Singh & Sandborn (2006); Sandborn (2013) |
| Maintainability | Decrease cost of ownership | Product | Virtue | Internal | Tortorella (2015) |
| Serviceability | Decrease cost of ownership | Product | Virtue | Internal | Dewhurst (1996) |
| Robustness (Taguchi Method) | Reduce production costs | Product | Virtue | Internal | Yu & Ishii (1998); Knoll & Vogel (2009) |
| End-Of-Life | Reduce environmental impact | Product | Lifecycle | External | Allenby & Graedel (1993) |
| Remanufacture (DfRem) | Reduce production costs; reduce barriers to innovation | Product | Lifecycle | Internal | Hatcher et al. (2011) |
| Packaging | Reduce production costs; reduce environmental impact | Product | Lifecycle | Internal | Dowlatshahi (1996) |
| Failure Modes (DFMEA) | Reduce failure rate | Product | Lifecycle | Internal | Cutuli et al. (2006) |
| Materials | Increase satisfaction of stated and implied needs; reduce production costs | Product | Lifecycle | Both | Karana et al. (2015) |
| Material Substitution | Increase resilience | Product | Virtue | Internal | Ljungberg (2005) |
| Modularity | Reduce production costs; reduce barriers to innovation | Product | Virtue | Internal | Erixon (1996) |
| Miniaturization | Reduce production costs; reduce barriers to innovation | Product | Virtue | Internal | Niedermayer et al. (2007) |
| Technical Merit | Increase competitive advantage | Product | Virtue | Internal | Murdoch & Wallace (1996) * |
| Affordances | Increase satisfaction of stated and implied needs | Product | Virtue | Internal | Maier & Fadel (2001) |
| Accessibility/ User Empowerment | Increase satisfaction of stated and implied needs | Product | Virtue | External | Ladner, R. E. (2015) |
| Lifecycle | Reduce lifecycle costs | System | Lifecycle | External | Chiu & Okusan (2010) |
| Supply Chain Management (DFSCM) | Reduce supply chain costs | System | Lifecycle | External | Lee & Sasser (1995) |
| Network | Reduce supply chain costs; Increase satisfaction of stated and implied needs | System | Lifecycle | External | Maltzman et al. (2005) |
| Transportability | Reduce supply chain costs | System | Lifecycle | External | Dowlatshahi (1999) |
| Logistics (DFL) | Reduce supply chain costs | System | Lifecycle | External | Mather (1992) |
| Storage and Distribution | Reduce supply chain costs | System | Lifecycle | External | Gopalakrishnan et al. (1996) |
| Mass Customization | Increase satisfaction of stated and implied needs; reduce production costs; reduce barriers to innovation | System | Lifecycle | External | Tseng & Jiao (1998) |
| Adaptability | Reduce lifecycle costs; reduce barriers to innovation | System | Virtue | Internal | Gu et al. (2004) |
| Lean Six Sigma | Reduce environmental impact | System and Ecosystem | Lifecycle | External | Jugulum & Samuel (2010) |
| Reverse Logistics (for Sustainability) | Reduce supply chain costs; reduce environmental impact | System and Ecosystem | Lifecycle | Both | Hosseini et al. (2015) |
| Electrostatic Discharge | Reduce environmental impact | Ecosystem | Lifecycle | External | Welsher et al. (1990) |
| Electromagnetic Compatibility | Reduce environmental impact | Ecosystem | Lifecycle | External | Dawson et al. (1996) * |
| Sustainability | Reduce environmental impact | Ecosystem | Virtue | External | Bhamra & Lofthouse (2007) |
| Environment (DFE) | Reduce environmental impact | Ecosystem | Virtue | External | Knight & Curtis (2002); O'Shea (2004) |
| Recycling | Reduce environmental impact | Ecosystem | Lifecycle | External | Gaustad et al. (2010) |
| Disposability | Reduce environmental impact | Ecosystem | Virtue | External | Blanchard et al. (1990) |
| Life Extension | Reduce environmental impact | Ecosystem | Lifecycle | External | Ljungberg (2005) |
| Energy Recovery & Substance Reduction | Reduce environmental impact | Ecosystem | Virtue | External | Fiksel & Wapman (1994); Ljungberg (2005) |
| Mood | Enhance affect | Ecosystem | Virtue | External | Desmet (2015) |
| Affect | Increase satisfaction of stated and implied needs | Ecosystem | Virtue | External | Shively & Balasubramanian (2003) |
| Inclusiveness | Expand market | Ecosystem | Virtue | External | Dong & Clarkson (2005) |
| Happiness | Increase emotional satisfaction | Ecosystem | Virtue | External | Hassenzahl et al. (2013) |
| Emergence | Increase satisfaction of stated and implied needs; reduce production costs; reduce barriers to innovation | Ecosystem | Virtue | External | Dogaru (2008) |

As noted by Meier & Fadel (2009), the development of DfX tools has historically been ad hoc, a reaction to the need for a product to have a certain quality attribute. In general, DfX design methods lack a theoretical basis, with the exception of axiomatic design. These authors attributed this criticism to Herbert Simon, who developed decision-making methods for design that were grounded in theory as early as the 1960's.

## Composition and Integration of Design for X (DfX) Techniques

Compared to the volume of papers that are focused on a single DfX, few researchers have explored using multiple DfX to optimize product design from a more holistic perspective, or articulating the composition of DfX in general to more easily develop new methods. This section is dedicated to the researchers that have emphasized these issues.

Huang & Mak (1997) identified that scholars at the time were engaged in the "search for a basic

DfX pattern, which can be used to explain how DfX works." They proposed a framework called the DfX Shell, a 7 step template for developing new DfX, which includes: requirement analysis, product modeling using Bill of Materials (BOM), process modeling, selecting performance measures, compiling DfX manuals, and verification of the method. The process modeling step is further decomposed into representing the business processes, representing resource requirements at each step of a process, identifying where consumption takes place, and specifying how resources are consumed by activities.

Bauer (2003) describes DfX as a system of knowledge, with the purpose of more effectively asking design questions and making decisions about tradeoffs. He proposed a hierarchical structure for the relationships between all DfX, categorized according to whether they pertained to the planning, concept design, or development phases. In addition, he proposed that there were three general classes of DfX: design for profit (immediate or future), design for resources (e.g. materials, environment, sustainability), and design for staff (which seemed to encompass any human aspect of a design problem).

Meerkamm & Koch (2007) added to Bauer's description of DfX, and suggested that each DfX could similarly be decomposed into a hierarchy to more effectively communicate the intent of the guidance within the DfX technique. Lindemann (2007) recognized that the application of DfX techniques was, in practice, mostly "chaotic" and recommended that teams apply network analysis, Design Structure Matrices (DSM) and Domain Mapping Matrices (DMM) to clarify the use of DfX at various stages, although it does not appear that other researchers have applied these recommendations yet.

Chiu & Okudan (2010) looked for similarities and patterns across elements of the DfX family. They recognized that there was a wide and diverse variety of DfX tools and techniques, and that each provided one or more of the following: 1) guidelines (often with measurable targets), 2) checklists, 3) metrics, 4) methods, and 5) mathematical models. Becker & Wits (2013) further deconstructed the architecture of the DfX tools, and identified three main components that are present in each one: the strategy declaration, the information type declaration, and the design task support method.

Nadadur & Parkinson (2013) linked human factors and ergonomics, a discipline that enables the design of products, services, tasks, and environments to serve the physical and cognitive requirements of users, to Design for Sustainability (DfS). By raising awareness of the diversity of human needs and the extent of human variability, especially in multinational contexts, their work suggests that ergonomics considerations should be integrated into *other* DfX rather than standing alone. Kuo et al. (2001) reviewed all of the DfX at the time and identified gaps. They recommended that future DfX research should incorporate findings from human factors engineering, human-technology interface studies, and the theory of learning to make systems more adaptive. Although intelligent systems and AI were recognized as powerful tools that could be useful in the selection of design alternatives, the widespread deployment of these systems was not noted as a potential driver for DfX research or use.

The range of design guidelines across DfX techniques was examined by Dombrowski et al. (2014), who concluded that cohesion between DfX could be achieved if they are applied in the context of Lean Design. This approach examines the product lifecycle from two perspectives: the *value* view, where the product is defined by the sum of the functions it offers and properties it has, and the *waste* view, which asserts that the product is the sum of all lifecycle processes (through end-of-life and beyond).

Keil & Lasch (2015) specifically focus on applying the family of DfX to process design and improvement, recognizing that DfX have traditionally been applied to new product development rather than innovation based on existing products. They apply Analytic Network Process (ANP), based on the Analytical Hierarchy Process (AHP) for multi-criteria decision making (MCDM), to illustrate the utility of this approach. This research suggests that the next generation of DfX should instead focus on the use of information and multi-criteria decision making that must support design, that is, a new family of "Decision Model for X" techniques. This supports the conclusion of Holt & Barnes (2010) who explain that research to ascertain the logical coherence between DfX and MCDM is needed: how can decision-making techniques benefit from the information that DfX provides?

## Quality Attributes for the IoT Ecosystem

Quality attributes are expressed by nonfunctional requirements that address the operation of a system rather than how well a system executes actions or behaviors. They "are just as critical as the functional features and user stories, as they assure the usability and effectiveness of the entire IoT ecosystem." (Fernandez & Pallis, 2014) Because IoT is dependent upon software and embedded systems, the eight quality attributes from ISO 25010 and fifteen quality attributes from the ISO 25012 data quality framework were used as the foundation for extracting quality attributes for the IoT ecosystem. (Rafique et al., 2012)

Croes & Hoepman (2015) presented a comprehensive examination of desired quality attributes for IoT components. Starting with the ISO 25010 guidance, they add the recommendations from an expert panel for the Internet of Things Architecture (IoT-A) institute. (Bassi et al., 2013) Integrating the quality attributes outlined by Fernandez & Pallis (2014) for healthcare IoT applications, removing duplicates, and consolidating the remaining attributes yields the lists that are shown in Figure 7. Croes & Hoepman (2015) make the very important point that depending what *kind* of IoT product you are designing, pairwise comparisons of the significance of the nonfunctional requirements will be different. As a result, AHP will be conducted for two product types (a wearable health sensor, and an enterprise-class package for home automation). The code to run the AHP will be provided so that the designer can consider other prioritizations of nonfunctional requirements during the design activities.

**Figure 7.** Quality attributes (nonfunctional requirements) for information-rich products.

| ***Product Criteria From ISO 25010 Rafique et al. (2012)*** | ***Product Criteria From Other Sources Croes & Hoepman (2015) Fernandez & Pallis (2014) Bassi et al. (2013)*** | ***Data-Based Criteria Based on ISO 25012 Rafique et al. (2012) Roy et al. (2014) Kiritsis (2010)*** |
|---|---|---|
| • functional suitability within intended environment<br>• testability and maintainability<br>• reliability and stability<br>• operability and accessibility<br>• performance efficiency<br>• transferability (between platforms)<br>• security and privacy | • evolvability (modularity, modifiability, extensibility, open source, compatibility)<br>• power efficiency and sustainability<br>• scalability<br>• safety<br>• robustness<br>• availability<br>• resiliency<br>• flexibility | • data **quality** (accuracy, precision, consistency, understandability, credibility, compliance, uncertainty)<br>• data **completeness** (availability, presence of metadata, including intent)<br>• data **lineage** (accessibility, traceability, provenance, confidentiality)<br>• data **portability** (can it be moved, can it change platforms) |

Other researchers establish the need for particular quality attributes more indirectly. Kiritsis (2010) defines an *intelligent product* which allows for "monitoring new parameters of the product and its environment along its whole [data] lifecycle." Roy et al. (2014) explored the applicability of data mining to data-driven decision making in manufacturing, and found that design intent, intelligent selection of materials, more effective assessment and selection of design alternatives, and enhancing inspection and post-sale support processes. They point out the uncertainty inherent in data collection on a large scale, and the challenges that may arise while attempting to integrate data from different areas of operations and at different scales.

# Results

With the nonfunctional requirements extracted, the analysis portion of this study involved two activities: 1) mapping the 50 DfX processes from Figure 4 to the customer satisfaction strategies within Golder et al.'s (2012) integrative quality framework, and 2) using the Analytical Hierarchy Process (AHP) to rank and prioritize DfX.

Linkages between DfX and customer satisfaction strategies are shown in Figure 8. Each of the strategies is product-centric and is sensitive to the difference between levels of functional requirement: “will” indicates a non-negotiable preference, while “should” is not nearly as strong, and “ideal” represents a preference that may not be required, but that defines the overall attractiveness of a product. For example, “Decrease ‘should’ expectation” means that the DfX should narrow the customer’s expectation of a product’s ability to satisfy a functional requirement, whereas “Decrease ‘should’ uncertainty” indicates that the DfX should increase the precision of the customer’s expectations. By connecting DfX to the perceptions and emotions of customers and stakeholders, we relate DfX not only to IoT but also to positive psychology (Radziwill, 2013) which is identified as an emerging research front in quality management.

**Figure 8.** The relationship of DfX techniques to strategies for improving customer satisfaction, with gaps.

| ***Strategy for Improving Customer Satisfaction*** | ***Relevant DfX*** | ***Production*** | ***Evaluation*** | ***Experience*** |
|---|---|---|---|---|
| Decrease “should” expectation | Design for Variety, Design for Mass Customization | | X | |
| Decrease “should” uncertainty | *None Available* | | X | |
| Increase “should” expectation | Design for Variety, Design for Mass Customization | | X | |
| Evoke positive emotion | Design for Quality, Design for Accessibility/User Empowerment, Design for Mood, Design for Affect, Design for Happiness | | | X |
| Evoke relevant emotion | Design for Quality, Design for Affordances, Design for Accessibility/User Empowerment, Design for Mood, Design for Affect, Design for Happiness | | | X |
| Increase attribute importance | Design for Variety, Design for Remanufacture, Design for Packaging, Design for Materials, Design for Material Substitution, Design for Modularity, Design for Affordances, Design for Mass Customization, Design for Adaptability, Design for Recycling, Design for Energy Recovery, Design for Mood | | X | |
| Decrease attribute importance | Design for Variety, Design for Remanufacture, Design for Packaging, Design for Materials, Design for Material Substitution, Design for Modularity, Design for Affordances, Design for Mass Customization, Design for Adaptability, Design for Recycling, Design for Energy Recovery, Design for Mood | | X | |
| Increase “ideal” uncertainty | *None Available* | | X | |
| Decrease “ideal” uncertainty | *None Available* | | X | |
| Move “ideal” expectation closer to perceived attribute | *None Available* | | X | |
| Change attribute design | Design for Manufacturing & Assembly, Design for Reusability, Design for Robustness, Design for Remanufacture, Design for Materials, Design for Material Substitution, Design for Affordances, Design for Energy Recovery, Design for Lifecycle, Design for Substance Reduction | X | | |
| Increase measurement knowledge and/or motivation to assess | Design for Accessibility/User Empowerment | | | X |
| Decrease measurement knowledge and/or motivation to assess | Design for Accessibility/User Empowerment | | | X |
| Increase expectation of what product "will" do | Design for Inclusiveness | | X | |
| Decrease expectation of what competitor "will" satisfy | *None Available* | | X | |
| Increase uncertainty regarding what product "will" do | *None Available* | | X | |
| Decrease uncertainty regarding what product "will" do | *None Available* | | X | |
| Improve process design specification through offline control (e.g. modeling and simulation) | Design for Cost, Design for Manufacturing & Assembly, Design for Variety, Design for Quality, Design for Six Sigma, Design for Reliability, Design for Obsolescence, Design for Robustness, Design for Materials, Design for Material Substitution, Design for Life Extension, Design for Substance Reduction | X | | |
| Improve process design specification through online methods (e.g. statistical process control) | Design for Cost, Design for Manufacturing & Assembly, Design for Six Sigma, Design for Reliability, Design for Maintainability, Design for Quality Manufacturability, Design for Failure Modes, Design for Sustainability, Design for Environment | X | | |
| Improve process design specification through inspection | Design for Cost, Design for Robustness, Design for Failure Modes | X | | |

The decision hierarchy was constructed with the primary goal at the top level ("Select DfX for IoT"), the Product Criteria from Figure 6 at the next level, Data Criteria from Figure 6 at the third level, and the 50 DfX processes at the lower level of Alternatives. One branch of the hierarchy is illustrated in Figure 9.

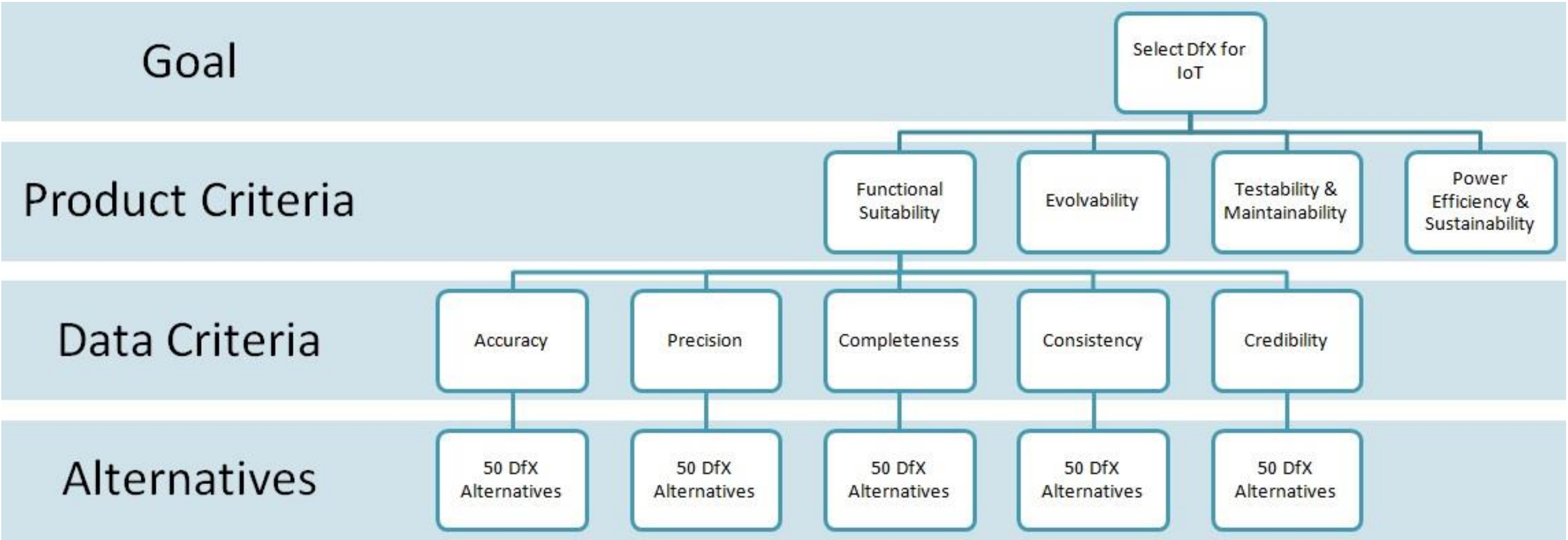

**Figure 9.** One partial branch of the decision hierarchy to select the best DfX for IoT.

The decision hierarchy for the initial run contained 15 product criteria, 4 data criteria, and all 50 DfX alternatives. The consistency, which indicates how internally consistent the judgments between criteria and subcriteria are, was in the desired range (< 10%) for all DfX and overall. From the initial run, 39 DfX and 9 product criteria were eliminated because their weights were much lower than the others. The weights in Figure 10 indicate the relative significance of each of the top 11 DfX methods with respect to the 6 product criteria that made the initial cut. Not surprisingly, evolvability, security and privacy, and power efficiency are the top considerations, and most of the DfX strongly support those concerns. Surprisingly, Design for Accessibility and Empowerment was highlighted, while Design for Network did not make the cut (despite the network-centric nature of the IoT).

**Figure 10. Global weightings associated with Top DfX alternatives.**

| | *Evolvability* | *Reliab. & Stability* | *Oper. & Access.* | *Perf. Eff.* | *Security & Privacy* | *Power Efficiency & Sust.* | *Safety* | *Overall* |
|---|---|---|---|---|---|---|---|---|
| **OVERALL** | 10% | 4.6% | 7.7% | 3.5% | 27.8% | 11.3% | 35% | 100% |
| Df Reliability | 1.3% | 0.5% | 1.6% | 0.4% | 3.2% | 1.4% | 4.1% | 12.5% |
| Df Accessibility/ Empowerment | 1.2% | 0.5% | 1.2% | 0.4% | 3.0% | 1.3% | 3.9% | 11.5% |
| Df Logistics | 1.1% | 0.5% | 0.9% | 0.4% | 3.1% | 1.3% | 4.0% | 11.4% |
| Df Robustness | 1.2% | 0.5% | 1.4% | 0.4% | 3.0% | 1.2% | 3.8% | 11.4% |
| Df Quality | 1.1% | 0.5% | 0.6% | 0.4% | 3.3% | 1.3% | 4.1% | 11.3% |
| Df Testability | 1.1% | 0.5% | 0.5% | 0.4% | 3.3% | 1.1% | 4.0% | 10.8% |
| Df Cost | 1.0% | 0.5% | 0.6% | 0.4% | 3.0% | 1.3% | 3.8% | 10.5% |
| Df Reusability | 1.0% | 0.5% | 0.5% | 0.4% | 2.9% | 1.2% | 3.7% | 10.2% |
| Df Mass Cust. | 1.0% | 0.5% | 0.5% | 0.4% | 2.9% | 1.2% | 3.7% | 10.2% |
| **CONSISTENCY** | 4.3% | 9.1% | 9.4% | 8.8% | 5.7% | 8.6% | 0.0% | 9.4% |

## Discussion and Conclusions

In this paper, mixed methods were employed to construct a gap analysis of "Design for X" techniques to help managers and researchers more easily select and apply these templates to design for the emerging Internet of Things (IoT). Using the Analytic Hierarchy Process (AHP), the most applicable DfX as-is were identified as the following: Design for Reliability, Design for Accessibility and User Empowerment, Design for Logistics, Design for Quality, Design for Robustness, Design for Testability, Design for Cost, Design for Reusability, Design for Mass Customization, Design for Adaptability, and Design for Environment. In addition, Dogaru's (2008) Design for Emergence has a strong theoretical basis, but significant work will be required before it is of practical use for designers.

This list should not be considered exhaustive, primarily due to the inherent limitations of the AHP method (where the accuracy of pairwise comparisons is highly sensitive to changes in technological capabilities and our attitudes about them). Additionally, this study only used “crisp” AHP where comparisons use only one synthesis of perspectives. Even though many perspectives were drawn in during the literature review, using only one synthesis could bias the results. Additionally, there are variations of AHP that accommodate multiple decision makers, variability in opinions on pairwise comparisons (stochastic AHP), and uncertainty in the ratings themselves (fuzzy AHP), and these could be valuable in generating a more universal list of applicable DfX for IoT.

New DfX methods must be accessible to non-experts, effectively applied to a new context, or must integrate a new, dynamic quality attribute that will be required for high-quality IoT products and processes. For the information-rich IoT, the development of these new DfX techniques should be grounded in theory, and follow established ("HMPR") guidelines for rigorous design science research in information systems (Hevner et al., 2004; Arnott & Pervan, 2012; Venable et al., 2012; Gregor & Hevner, 2013; Gill & Hevner, 2013) as well as guidance on agile frameworks (Benton & Radziwill, 2011) and what constitutes a DfX. (Huang & Mak, 1997)

Furthermore, the gaps in the existing collection of DfX call out a fundamental underlying assumption of design to date: that we do not have access to the customer or stakeholder after the point of sale, and so cannot shape or validate their opinions about our products' functional or nonfunctional attributes over time. As a dynamic, aware, information-rich context for generating new information and developing ongoing relationships, the IoT ecosystem *makes this assumption obsolete.* New DfX can assume *panpsychism*; that is, how do you structure your design methods when the designed objects and the environment within which they are situated are themselves *aware* and you have access to what they learn about themselves over time? (Karman, 2012; Dehaene et al., 2014; Wittgenstein, 2015) This could change the fundamental foundations of design thinking.

The results from a comprehensive literature review, a map of the corpus of DfX to customer satisfaction strategies from an integrative quality framework, and the classification from Bauer (2003) were joined to illuminate gaps. Each of these gaps arises due to the richness of the dynamic information context that is provided by the IoT. As a result, the next generation of DfX techniques should include:

- **Cost-centric DfX**: to make quicker, better, and smarter decisions with IoT data; to effectively deploy data management strategies; to achieve traceability and provenance of data sources and computational evolution of data products to reduce lifecycle costs.
- **Resource-centric DfX**: to preserve battery life and harvest energy; reduce the amount of data transfer over the network/reduce the load on the cloud; to integrate considerations for spectrum management and radio frequency interference (RFI) mitigation; to increase and enable learning (from the resource perspective); to manage emergent behavior of the IoT component; or DfX in the context of Lean Design to better address resource constraints.

- **Human-centric DfX**: to protect the privacy of individual users (e.g. via differential privacy); to preserve agency with respect to the goal-setting behavior of intelligent agents and support notion of consent; to acknowledge and leverage natural human variation (including cognitive variation such as neurodiversity); to increase and enable learning (from the individual perspective); to make time-sensitive decisions at various levels of analysis with imperfect information and uncertainty; to expand the designer's perception of *experience* beyond a user's interaction with the product's interfaces.

Quality managers, and engineering managers focused on designing quality into systems, will be the audience for new DfX. In the most recent Future of Quality report from the American Society for Quality (ASQ), Snee and Hoerl (2015) identify some areas of focus for quality managers as IoT expands. They include the need to apply an increasingly holistic improvement approach, to leverage Big Data to solve problems previously considered unsolvable, and to better address the positive potentials of human variability. Because the DfX techniques naturally provides a holistic basis for improvement at the design stage, further development of DfX should also include designing in the ability to benefit from Big Data, and drawing from human factors engineering to provide guidance for effectively leveraging human potential by embracing variation.

The IoT is not merely a network, but an emergent and continuous process through which people and machines co-create knowledge. In the IoT, the process of information production and synthesis *is* the product, the producer is also the consumer, and design support tools must provide enhanced decision support. (Holt & Barnes, 2010; Keil & Lasch, 2015) A new family of “Decision Model for X” techniques, to help designers use the *dynamic*, rich IoT information streams to improve managerial decision making, is thus needed to fill the gap. What we have previously recognized as continuous improvement frameworks for networks of interconnected processes, such as the Malcolm Baldrige National Quality Award (DeJong, 2009) and Capability Maturity Model for Integration (CMMI Team, 2002) have the potential to become design templates for products in the IoT because of their utility as decision support systems. The *people* in those continuous improvement frameworks become the *entities* in the IoT, because the roles of human and machine are blurred. As a result, an important emerging research opportunity lies in translating process improvement frameworks into the language of IoT products.